\documentclass[pra, showpacs, preprint, floatfix]{revtex4}
\usepackage{graphicx}
\usepackage{epstopdf}
\usepackage{amsmath, amsfonts, amssymb, bm}

\begin{document}

\title{Phase-dependent quantum interferences with three-level artificial atoms}
\author{Victor \surname{Ceban}}
\email{victor.ceban@phys.asm.md}
\affiliation{Institute of Applied Physics, Academy of Sciences of Moldova, 
Academiei str. 5, MD-2028 Chi\c{s}in\u{a}u, Moldova}

\begin{abstract}
The phase dependence of the cavity quantum dynamics in a driven equidistant three-level ladder-type system found in a quantum well structure with perpendicular transition dipoles is investigated in the good cavity limit. The pumping laser phases are directly transferred to the superposed amplitudes of the cavity-quantum-well interaction. Their phase difference may be tuned in order to obtain destructive quantum interferences. Therefore, the cavity field vanishes although the emitter continues to be pumped. 
\end{abstract}

\maketitle

\section{Introduction}
The confinement of quantum systems in a specific superposition of states may lead to various quantum processes. In the realm of quantum optics, a particular interest is focused on this type of phenomena, namely,  quantum interference phenomena allow to explain and observe various quantum effects \cite{aga74, scu97, koc88, bol91, kei99}, while etanglement processes \cite{mar03, isa13,isa15,isa16} and shape-preserving localized light structures \cite{mih12, mih13} play a major role for the quantum computation and communication. A powerful tool in the control and manipulation of these effects originates from an additional degree of freedom of the system given by its phase dependence. For example, quantum interference effects influence the collective fluorescence of a driven sample of emitters, which becomes sensitive to phase dependence. Thus, the phase difference of the two lasers pumping a collection of three-level emitters may decrease and cancel its fluorescence when quantum interferences appear from a coherently driven source \cite{mac03}. The superflourescent behaviour of a sample of four-level emitters is modified by the vacuum induced quantum interferences and may be enhanced by varying the phase difference of the pumping lasers \cite{mac07a}. Moreover, for a well-chosen phase the sample may be trapped in its excited state and thus decoupled from the surrounding environment. The phase dependent complete or partial cancellation of the spontaneous emission is reached when a single four-level emitter is considered \cite{pas97}. The spontaneous emission properties may also be controlled via the phase difference of the pumping laser and a squeezed surrounding reservoir for a three-level ladder-type emitter \cite{mac07b}. In a different scenario, phase dependent systems may be used to study the phase itself, e.g. , the carrier-envelope phase of a few-cycle laser pulse may be determined via the behaviour of the populations of a qubit system \cite{xie09}. 

A more challenging goal has been the realization of quantum effects in systems made of artificial atoms such as quantum wells (QWs), as these systems possess additional degrees of freedom, which leads to stronger decoherent phenomena \cite{sad98}. The particular interest in this type of artificial atoms for the current realm is the possibility to tailor their energetic states via the layer thicknesses and materials used for the QW \cite{ros91}. Quantum interference phenomena as gain without inversion have been experimentally obtained for pumped three-level ladder-type coupled triple wells \cite{fro06}, while electromagnetically induced transparency has been observed in three-level QW systems with $\Lambda$-type transitions \cite{phi02} as well as ladder-type intersubband transitions \cite{dyn05, ser00}. A direct detection of ac Stark splitting, i.e. , dressed-state splitting, has been experimentally achieved in \cite{dyn05} for $\Xi$-type QWs. This type of QWs is particularly interesting as it may be engineered as an equidistant three-level emitter \cite{sad98, ros91}, an emitter difficult to implement with real atoms.

In this paper, a pumped ladder-type three-level QW placed in a cavity is investigated. The QW architecture has equidistant energy levels and orthogonal transition dipoles. Each transition is resonantly driven by lasers with different phases. The energy level distribution allows the optical cavity to couple with each of the QW transitions. Under the laser driving, the QW exciton is prepared in a superposition of states, which leads to quantum interference of the indistinguishable amplitudes of the cavity interaction with the different exciton transitions. Strong destructive interferences may be achieved if the cavity is tuned to the most or less energetic dressed-state transition of the pumped QW. Therefore, the cavity field may be emptied for a well-chosen laser phase difference as the laser phases are transferred to the interactional amplitudes.  In this case, the pumped QW spontaneously decays in all directions except the cavity. Furthermore, this behaviour of the interfering QW-cavity system is associated with a quantum switch, where the income laser signals may switch the cavity field on and off by varying their phase difference.

This article is organized as follows. In Sec. 2 the studied model is described, one presents the system Hamiltonian, the applied approximations and the master equation solving technique. The results on the quantum interferences effect are discussed in Sec. 3. The summary is given in Sec. 4.

\section{The model}

The model consists of a three-level equidistant ladder-type QW placed in an optical cavity. The QW is driven by two intense lasers and has perpendicular transition dipoles, which allows to set each laser to pump a separate transition. The QW is described by its bare-states $\vert i \rangle , \lbrace i=1,2,3 \rbrace$ and their corresponding energies $\hbar \omega_{i}$. The atomic operators are defined as $S_{ij} = \vert i \rangle \langle j \vert$, $\{i,j = 1,2,3 \}$ and obey the commutation rule $[S_{\alpha,\beta},S_{\beta', \alpha'}] = \delta_{\beta ,\beta'}S_{\alpha,\alpha'}
-\delta_{\alpha',\alpha}S_{\beta',\beta}$. The most energetic level $\vert 3 \rangle$ may spontaneously decay to the intermediate level $\vert 2 \rangle$  with a rate $\gamma_{2}$, while the last one decays to the ground level  $\vert 1 \rangle$  with a rate $\gamma_{1}$. The laser pumping of the QW is expressed by semi-classical interactions with Rabi frequency $\Omega_{1}$ ($\Omega_{2}$) corresponding to the laser of frequency $\omega_{L1}$ ($\omega_{L2}$) and phase $\phi_{1}$ ($\phi_{2}$) driving the lower (upper) transition. The QW-cavity quantum interaction is described by the coupling constant $g_{1}$ ($g_{2}$) corresponding to the interaction of the optical resonator with the lower (upper) QW transition. The cavity field is defined by its frequency $\omega_{c}$ and the bosonic creation (annihilation) operators $a^{\dagger}$ ($a$) that commute as $[a, a^{\dagger}] = 1$. The cavity is dumped by a vacuum reservoir at a rate $\kappa$. The system Hamiltonian is defined as:
\begin{equation}\label{Htot}\begin{aligned}
H =& \hbar\omega_{c} a^{\dagger}a + \hbar \sum^{3}_{i=1} \omega_{i}S_{ii} + i \hbar g_{1} (a^{\dagger} S_{12} - S_{21} a )+ i\hbar g_{2} (a^{\dagger} S_{23} - S_{32} a )  \\ 
&+ \hbar \Omega_{1}(S_{21}e^{-i (\omega_{L1} t + \phi_{1})} + S_{12}e^{i (\omega_{L1} t + \phi_{1})}) \\
&+ \hbar \Omega_{2}(S_{32}e^{-i (\omega_{L2} t + \phi_{2})} + S_{23}e^{i (\omega_{L2} t + \phi_{2})}).
\end{aligned}\end{equation}
where the first two terms are the free cavity and QW terms, the next two terms represent the QW-laser semi-classical interaction, while the last two terms describe the QW-cavity quantum interaction. The system dynamics is described by the master equation of the density operator $\rho$, namely:
\begin{equation}\label{Meq}\begin{split}
\frac{\partial \rho}{\partial t} &= - \frac{i}{\hbar} [ H, \rho ] + \frac{\kappa}{2} \mathcal{L} (a) + \frac{\gamma_{2}}{2} 
\mathcal{L} (S_{23})+ \frac{\gamma_{1}}{2} \mathcal{L} (S_{12}),  \\
\end{split}\end{equation}
where the Liouville superoperator is defined as $\mathcal{L} (\mathcal{O}) = 2 \mathcal{O} \rho \mathcal{O}^{\dagger} - \mathcal{O}^{\dagger} \mathcal{O} \rho - \rho \mathcal{O}^{\dagger} \mathcal{O}$ for a given operator $\mathcal{O}$. The second term of the equation describes the cavity damping, while the last two terms represents the QW spontaneous emission.

In the interaction picture, the Hamiltonian is brought to an easy diagonalizable form of the QW-lasers subsystem terms and is defined as: 
\begin{equation}\label{Htran}\begin{aligned}
H =& \hbar(\omega_{c} - \omega_{L}) a^{\dagger}a + \hbar \Omega_{1}(S_{21} + S_{12}) +\hbar \Omega_{2}(S_{32} + S_{23})  \\ 
&+ i \hbar g_{1} (a^{\dagger} S_{12} e^{-i \phi_{1}} - e^{i \phi_{1}} S_{21} a ) + i\hbar g_{2} (a^{\dagger} S_{23} e^{-i \phi_{2}} - e^{i \phi_{2}} S_{32} a ).
\end{aligned}\end{equation}
Here, the lasers are considered to be resonant with the QW transitions and therefore $\omega_{L1} = \omega_{L2} = \omega_{L}$. Next, one adopts the semi-classical dressed-state transformation according to the dynamical Stark splitting effect of the QW under the laser pumping \cite{dyn05}. In analogy with the Mollow triplet of a two-level emitter, the fluorescence spectra of the driven QW possess sidebands that are symmetrical to the central bar-state frequency peak. However, in the case of the equidistant three-level emitter one has four degenerate sidebands due to its degenerate bare-state central peak. The new Hermitian base is defined considering the pumped QW subsystem eigenfunctions. The new atomic wavefunction basis vectors, i.e., the dressed-states, are defined as \cite{ceb16}:
\begin{equation}\label{dsb}\begin{split}
\vert 1 \rangle &= -\frac{1}{\sqrt{2}} \cos{\theta}\, \vert -\rangle - \sin{\theta} \, \vert 0 \rangle + \frac{1}{\sqrt{2}} \cos{\theta} \, 
\vert + \rangle , \\ 
\vert 2 \rangle &=  \frac{1}{\sqrt{2}} \, \vert -\rangle  + \frac{1}{\sqrt{2}} \, \vert + \rangle ,  \\ 
\vert 3 \rangle &= -\frac{1}{\sqrt{2}} \sin{\theta} \, \vert -\rangle + \cos{\theta} \, \vert 0 \rangle + \frac{1}{\sqrt{2}} \sin{\theta}\, 
\vert + \rangle, 
\end{split}\end{equation}
where $\theta=\tan^{-1}(\Omega_{2}/\Omega_{1})$, $\Omega=\sqrt{\Omega_{1}^{2}+ \Omega_{2}^{2}}$. Once, the dressed-state transformation is applied, one tunes the cavity in resonance with the sideband transitions of the dressed-QW, i.e., $\omega_{c} = \omega_{L} \pm \Omega$ or $\omega_{c} = \omega_{L} \pm 2 \Omega$. At the resonance, one may apply a secular approximation to the Hamiltonian expressed in the interaction picture, where one may keep only the resonant terms as long as $g_{1,2} \ll \Omega$. In what follows, the cavity is set in resonance to the most energetic sideband, but note that the further discussions and results are also valid for the case when the cavity is tuned to the less energetic transition, where a similar behaviour is observed. The Hamiltonian within the secular approximation is brought to the form:
\begin{equation}\label{Hfin}\begin{split}
H &= ig(a^{\dagger}R_{-+}e^{i\psi}  - e^{-i\psi}R_{+-}a) ,\\
\end{split}\end{equation}
where $g=(g_{2}e^{-i\phi_{2}} \sin{\theta} - g_{1}e^{-i\phi_{1}} \cos{\theta} )/2$ and $\psi = \text{arg}⁡(g)$. The new set of atomic dressed-state operators is defined as $R_{ij}=\vert i \rangle \langle j \vert$, $\{i,j\} \in \{-, 0, +\}$ and $R_{z} = R_{++} - R_{--}$. The new operators obey the same commutation relations as the previous ones. The master equation is defined in the new basis as: 
\begin{equation}\label{MeqDS}\begin{split}
\frac{\partial \rho}{\partial t} &= - \frac{i}{\hbar} [ H, \rho ]+ \frac{\kappa}{2} \mathcal{L} (a) +  \gamma_{a} (\mathcal{L} (R_{-0}) +\mathcal{L} (R_{+0}))  \\
&+ \gamma_{b} (\mathcal{L} (R_{0-}) +\mathcal{L} (R_{0+}))+ \gamma_{c} \left( \mathcal{L} (R_{z}) +\mathcal{L} (R_{+-}) + \mathcal{L} (R_{-+}))\right) , 
\end{split}\end{equation}
where $\gamma_{a}=\gamma_{2} (\cos^{2}\theta)/4$, $\gamma_{b}=\gamma_{1} (\sin^{2}\theta)/4$ and $\gamma_{c}=(\gamma_{2}  \sin^{2}\theta + \gamma_{1} \cos^{2}\theta )/8$ \cite{ceb16}. Similarly to \cite{cio13}, a secular approximation was applied on the QW spontaneous emission fast-rotating terms, in order to obtain the equation (\ref{MeqDS}). Note that the laser is considered enough intense to satisfy the secular approximation condition $\gamma_{1,2} \ll \Omega$.

The master equation is numerically solved via projecting it in the system state basis \cite{qua93}. The solving technique was adapted to the case when a three-level emitter and phase dependent lasers are considered. A first projection in the QW dressed-states leads to a system of linear differential coupled equations defined by the variables: $\rho^{(0)} = \rho_{--}+\rho_{00}+\rho_{++}$, \, $\rho^{(1)}= \rho_{++}+\rho_{--}$, \, $ \rho^{(2)}= \rho_{++}-\rho_{--}$, \, 
$\rho^{(3)} = (a^{\dagger} \rho_{+-}e^{i\psi}+e^{-i\psi}\rho_{-+}a)/2$, and $\rho^{(4)} = (\rho_{+-} a^{\dagger}e^{i\psi} +e^{-i\psi}a\rho_{-+})/2 $, where 
$\rho_{ij}=\langle i \vert \rho \vert j \rangle$, $\{i,j \in -, 0, +\}$ are the QW reduced density matrix elements. The equations are next projected in the cavity field Fock states basis $\lbrace\vert n\rangle, n\in \mathcal{N}\rbrace$, leading to the following set of equations:
\begin{equation}\label{syst}\begin{split}
\dot{P}^{(0)}_{n}  =& -2 \vert g\vert (P_{n}^{(4)}-P_{n}^{(3)}) + \kappa (n+1) P_{n+1}^{(0)} - \kappa n P_{n}^{(0)} ,  \\
\dot{P}^{(1)}_{n} =& -2 \vert g\vert (P_{n}^{(4)}-P_{n}^{(3)}) + \kappa (n+1) P_{n+1}^{(1)} -( \kappa n +\alpha/2) P_{n}^{(1)} + \gamma_{2} \cos^{2}{\theta} P_{n}^{(0)},  \\
\dot{P}^{(2)}_{n} =& -2 \vert g\vert (P_{n}^{(4)}+P_{n}^{(3)}) + \kappa (n+1) P_{n+1}^{(2)} - (\kappa n + \beta/2) P_{n}^{(2)} ,  \\ 
\dot{P}^{(3)}_{n} =& \vert g\vert n (P_{n-1}^{(1)} - P_{n}^{(1)} +P_{n-1}^{(2)} +P_{n}^{(2)})/2 - \kappa P_{n}^{(4)} + \kappa (n+1) P_{n+1}^{(3)} \\
&- (\kappa (n-1/2) + \zeta) P_{n}^{(3)}  ,  \\ 
\dot{P}^{(4)}_{n} =& \vert g\vert (n+1)(P_{n+1}^{(2)} + P_{n}^{(2)} - P_{n+1}^{(1)} + P_{n}^{(1)})/2 + \kappa (n+1) P_{n+1}^{(4)} \\
&- (\kappa (n+1/2) + \zeta) P_{n}^{(4)}, 
\end{split}\end{equation}
where
$\zeta = [\gamma_{1} (2 + \cos^{2}{\theta}) + 3 \gamma_{2} \sin^{2}{\theta}]/4$, \, $\alpha = \gamma_{1}\sin^{2}{\theta} + 2 \gamma_{2}\cos^{2}{\theta}$, \, 
$\beta = \gamma_{1} + \gamma_{2} \sin^{2}{\theta}$, 
and $P_{n}^{(i)}= \langle n\vert \rho^{(i)} \vert n\rangle $. 

This system of equations (\ref{syst}) is numerically solved, considering the probability conservation of the density matrix elements, i.e. , $\text{Tr}[\rho]=1$, and their asymptotic behaviour that allows the system to be truncated at a certain maximum $n_{max}$ of considered Fock states. The parameters of interest are estimated from the system variables and will be presented and discussed in the next Section. One observes the cavity behaviour via the mean photon number $\langle n \rangle$ and the second-order photon-photon correlation function $g^{(2)}(0)$ defined by the diagonal elements of the QW's reduced density matrix, deduced from the system (\ref{syst}) as follows:
\begin{equation}\label{nn}\begin{split}
\langle n \rangle = \langle a^{\dagger} a \rangle  =  \sum_{n=0}^{\infty} i P_{i}^{(0)} \simeq \sum_{n=0}^{n_{max}} i P_{i}^{(0)} ,
\end{split}\end{equation}
\begin{equation}\label{g2}\begin{split}
g^{(2)}(0) = \frac{\langle a^{\dagger} a^{\dagger} a a \rangle}{\langle a^{\dagger} a \rangle^{2}} 
= \frac{1}{\langle n \rangle^{2}} \sum_{n=0}^{\infty} i (i-1) P_{i}^{(0)} \simeq \frac{1}{\langle n \rangle^{2}} \sum_{n=0}^{n_{max}} i (i-1) P_{i}^{(0)}. 
\end{split}\end{equation}

\section{Results and discussions}

\begin{figure}[t]
\centering
\includegraphics[width=0.8\textwidth]{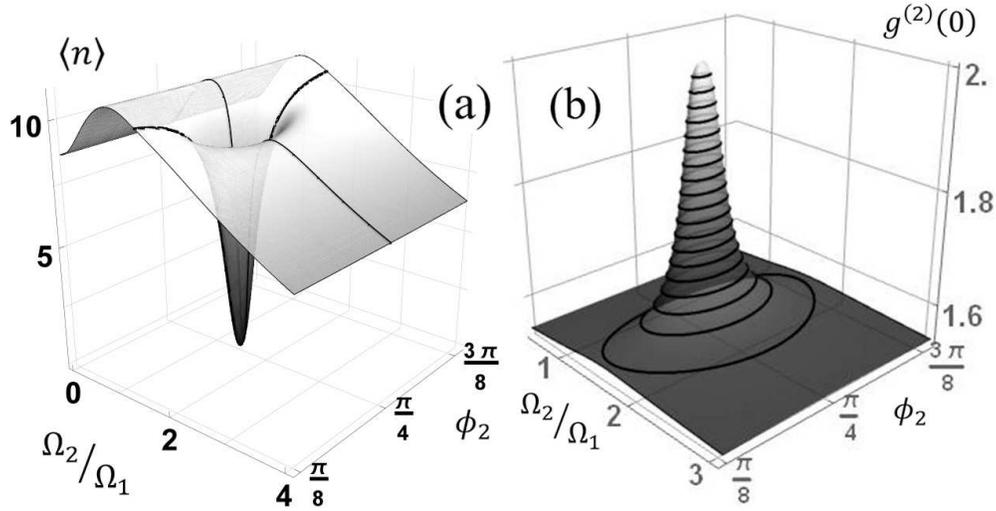}
\caption{(a): The cavity mean photon number $\langle n \rangle$ and (b): the second-order photon-photon correlation function $g^{(2)}(0)$ as functions of the laser phase $\phi_{2}$ and Rabi frequencies ratio $\Omega_{2}/\Omega_{1}$. Here $g_{1}/\gamma_{1} =6$, $g_{2}/\gamma_{1} =4$, $\gamma_{2}/\gamma_{1} =2$, $\kappa/\gamma=10^{-3}$, and $\phi_{1} = \pi /4$.}
\label{pic1}
\end{figure}

The cavity field behaviour shows a good evidence of quantum interferences, as presented in Fig. \ref{pic1}. For a certain configuration of laser phases and Rabi frequencies ratio, the mean photon number is strongly decreased down to the zero value. This minimum describes a complete cancellation of the cavity field and corresponds to the case when the two indistinguishable amplitudes of the QW-cavity interaction are equal and in-phase. Therefore, when the cavity interacts equally with both of the QW transitions, the interaction amplitudes cancel each other due to their destructive superposition.  This destructive quantum interference effect is also reflected in the behaviour of the field second-order correlation function describing the photon distribution. When the cavity mean photon number is cancelled, $g^{(2)}(0) \rightarrow 2$, asymptotically describing a thermal distribution. The cavity is in equilibrium with the surrounding electromagnetic vacuum, when maximum interference effect is reached. 

The phase difference of the input lasers plays a crucial role in the control of the quantum interference. The interaction amplitudes phases are related to the laser phases as suggested by the expression of the coupling constant $g$ of the Hamiltonian form of equation (5) at cavity-QW resonance and within the secular approximation. Therefore, a destructive superposition is obtained when the interaction amplitudes are in-phase, i.e., $\phi_{2} = \phi_{1} + 2 \pi m, m \in \mathcal{Z}$, as shown in Fig. \ref{pic1}. At this condition, the system behaves similarly to the case when no laser phase was considered \cite{ceb16}, where the field cancels simply for $g_{1}/g_{2} = \Omega_{2}/\Omega_{1}$.

The possibility to control and turn-off the cavity field via quantum interferences suggests a potential application of the studied QW-cavity system for quantum network circuits \cite{kim08, ion15}. The model is sensitive to phase and intensity variations of the input lasers and acts as quantum switch, where the cavity field is turned on or off. Both input parameters are largely confined in experimental conditions. Moreover, artificial-atom-based systems could be relevant candidates for on-chip quantum circuits \cite{hou12}.  

\section{Summary}

The model of a pumped equidistant three-level ladder-type quantum well placed in an optical cavity has been investigated in the good cavity limit. The emitter has perpendicular transition dipoles and the cavity couples to both of the QW transitions. Two intense lasers with different phases are used to  resonantly drive the emitter and each laser couples semi-classically to a different transition. It has been shown that the laser phases are transferred to the QW-cavity interaction amplitudes. Therefore, the superposition of the indistinguishable amplitudes is phase dependent, so that the resulting destructive quantum interferences effect on the cavity field becomes sensitive to the phase difference of the input lasers.

\section*{Acknowledgement}
The author is thankful to M. A. Macovei for fruitful discussions related to this study. He acknowledges the financial support from the Academy of Sciences of Moldova via grant No. 15.817.02.09F.

\end{document}